\documentclass{aa}
\usepackage{graphicx}
\usepackage[varg]{txfonts}
\usepackage{longtable}
\usepackage{lscape}
\usepackage{color}

\begin{document}

 \title{Variability in stellar granulation and convective blueshift with spectral type and magnetic activity. II. From young to old main-sequence K-G-F stars\thanks{Table A.1 is only available in electronic form at the CDS via anonymous ftp to cdsarc.u-strasbg.fr (130.79.128.5) or via http://cdsarc.u-strasbg.fr/viz-bin/qcat?J/A+A/} }

   \titlerunning{Variability in stellar granulation and convective blueshift }

   \author{N. Meunier \inst{1}, L. Mignon \inst {1}, A.-M. Lagrange \inst{1}
          }
   \authorrunning{Meunier et al.}

   \institute{
Univ. Grenoble Alpes, CNRS, IPAG, F-38000 Grenoble, France\\
  \email{nadege.meunier@univ-grenoble-alpes.fr}
             }

\offprints{N. Meunier}

   \date{Received 24 April 2017 ; Accepted 29 August 2017}

\abstract{The inhibition of small-scale convection in the Sun dominates the long-term radial velocity (RV) variability: it therefore has a critical effect on light exoplanet detectability using RV techniques.}
{We here extend our previous analysis of stellar convective blueshift and its dependence on magnetic activity to a larger sample of stars in order to extend the T$_{\rm eff}$ range, to study the impact of other stellar properties, and finally to improve the comparison between observed RV jitter and expected RV variations.  }
{We estimate a differential velocity shift for Fe and Ti lines of different depths and derive an absolute convective blueshift using the Sun as a reference for a sample of 360 F7-K4 stars with different properties (age, Teff, metallicity). }
{We confirm the strong variation in convective blueshift with Teff and its dependence on  (as shown in the line list in Paper I) activity level. Although we do not observe a significant effect of age or cyclic activity, stars with a higher metallicity tend to have a lower convective blueshift, with a larger effect than expected from numerical simulations. Finally, we estimate that for 71\% of the stars in our sample the RV and LogR'{\rm HK} variations are compatible with the effect of activity on convection, as observed in the solar case, while for the other stars, other sources (such as binarity or companions) must be invoked to explain the large RV variations. We also confirm a relationship between LogR'{\rm HK} and metallicity, which may affect discussions of the possible relationship between metallicity and exoplanets, as RV surveys are biased toward low LogR'{\rm HK} and possibly toward high-metallicity stars. }
{We conclude that activity and metallicity strongly affect the small-scale convection levels in stars in the F7-K4 range, with a lower amplitude for the lower mass stars and a larger amplitude for low-metallicity stars.  }

\keywords{Physical data and processes: convection -- Techniques: radial velocities  -- Stars: magnetic field -- Stars: activity  -- 
Stars: solar-type -- Sun: granulation} 

\maketitle

\section{Introduction}

\cite{meunier17}, hereafter Paper I, measured the amplitude of the convective blueshift for a large
sample of stars. This covered G and K main-sequence stars for six spectral types, K2, K0, G8, G5, G2, and G0. The authors observed a strong decrease toward low-mass stars. Furthermore, we showed for the first time a significant dependence of the convective blueshift on the activity level. We found that the attenuation factor of the convective blueshift in solar-like plages (due to a higher concentration of the magnetic field) seems to follow a constant law as a function of spectral type. These results are crucial to characterize the effect of stellar activity on radial velocity (hereafter RV) measurements more
precisely, and following from this, the effect on exoplanet detectability \cite[][]{meunier10a}. In particular, the results enable us to extrapolate the simulations made by \cite{borgniet15} to stars other than the Sun. 

In this new study, we extend our sample by more than a factor of two (leading to 360 stars) with three main objectives: i) extension toward higher-mass stars (up to F7 stars)  and lower-mass stars (down to K4 stars) to extend the validity domain of our conclusions for both the dependence on spectral type and activity level; ii) study the dependence of the convective blueshift on other stellar properties, such as the type of variability (cyclic behavior or more stochastic variability), on age, and on metallicity, especially for solar-like stars, of which we have been able to select a large sample; iii) improvement of the comparison between the RV variability due to the attenuation of the convective blueshift with observations, based on a careful analysis of the relationship between RV and LogR'$_{\rm HK}$ variations in our sample, from which we derive some statistics. The sample studied in Paper I was restricted to relatively old main-sequence stars. However, we know that the activity patterns of younger stars are different from the Sun, with a stronger influence caused by spots over plages \cite[e.g.,][]{lockwood07}. Plages are much more common in these stars than in the Sun, however, as shown by the much larger LogR'$_{\rm HK}$  index, and it is not known whether their properties are different from solar properties. Moreover, numerical simulations have shown an effect of metallicity on convection
that is due to the differences in opacities, although with some unclear pattern \cite[][]{magic13,magic14,allendeprieto13,tremblay13}. 


As in Paper I, our analysis is based on the estimation of the differential velocity shift (i.e., computed over a set of lines of different fluxes), as previously done for the Sun \cite[][]{dravins81,hamilton99} and other stars  \cite[][]{gray82,dravins87b,dravins99,landstreet07,allendeprieto02,gray09,meunier17}. This is then converted into an absolute convective blueshift using the Sun as a reference, as in Paper I.  
The absolute convective blueshift varies from one line to the other and also depends on how it was measured. In the following, the reference to an absolute convective blueshift corresponds to a single value (see in particular Sect.~2.2) associated with a certain set of lines, following the procedure described in Paper I. This therefore needs to be kept in mind when attempting to compare absolute values with other works.
We focus our analysis on long-term variability rather than on variability at the rotational timescale, as has been done by \cite{suarez17}, and we also study a much larger sample.  
We note we focus here on the line shifts and not on the asymmetries of the lines that are due to convection or other effects such as rotation \cite[][]{gray86}.

The outline of the paper is the following. In Sect.~2 we describe our sample, and then summarize our method \cite[we refer to][Paper I, for more details]{meunier17}. In Sect.~3 we analyze the dependence of the  convective blueshift on spectral type (using B-V and Teff), on the activity level, on the type of variability, on age, and on metallicity. In Sect.~4 we derive statistical information on how the RV variations correlate with activity variations. This information allows us to estimate the effect that the inhibited convection blueshift has on stars other than the Sun, for which we know it is a dominating factor. Finally, we conclude in Sect.~5.  


\section{Data analysis}

\subsection{Star sample}

\begin{figure} 
\includegraphics{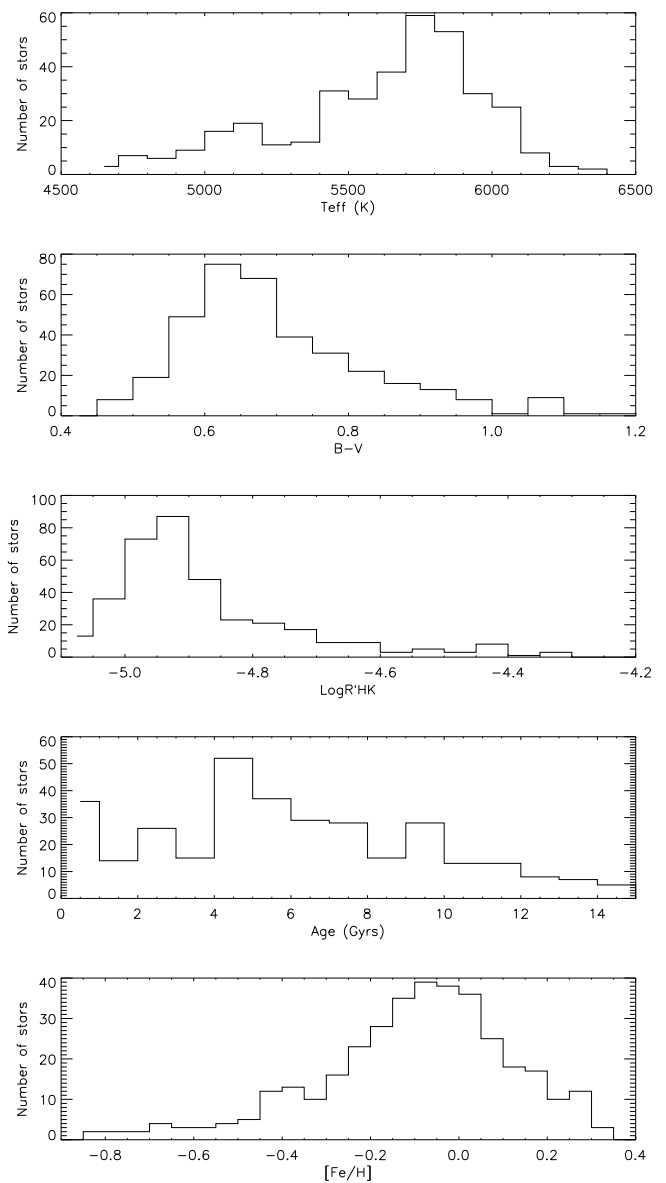}
\caption{
{\it First panel}: histogram of Teff in our sample.
{\it Second panel}: same for B-V.
{\it Third panel}: same for LogR'$_{\rm HK}$. 
{\it Fourth panel}: same for ages.
{\it Fifth panel}: same for [Fe/H].
}
\label{sample}
\end{figure}

\begin{table*}
\caption{Sample origin}
\label{tab_sample}
\begin{center}
\renewcommand{\footnoterule}{}  
\begin{tabular}{llll}
\hline
\# & Reference & Type of survey & Number of stars \\ \hline
1 & \cite{sousa08} & F-G-K stars, old, not very active, exoplanet survey & 294 \\
2 & \cite{ramirez14} & Solar twins, old and young stars & 64 \\
3 & \cite{marsden14} & Solar-type stars, old and young (Bcool survey) & 31 \\
4 & \cite{datson14} & Solar twins & 5 \\
5 & \cite{borgniet17} & F stars, old stars, exoplanet survey & 6 \\
6 & \cite{lagrange13} & Young stars, exoplanet survey & 8 \\
7 & \cite{gray15} & Young solar twins & 4 \\
\hline
\end{tabular}
\end{center}
\tablefoot{Number of stars retrieved from these surveys to build our sample. }
\end{table*}

Our sample includes 360 stars with spectral types from F7 to K4 (spectral types and B-V values were retrieved from the Simbad database at the CDS, https://simbad.u-strasbg.fr/simbad/), observed with HARPS by several groups: \cite{sousa08}, \cite{ramirez14}, \cite{marsden14}, \cite{datson14}, \cite{borgniet17}, \cite{lagrange13}, and \cite{gray15}. The spectra are available in the ESO archives, and we selected only spectra with an average signal-to-noise ratio above 100, leading to a total of 19510 spectra (i.e., 54 spectra on average per star, a median value of 25, and a number of spectra per star from 2 to 1230). HARPS spectra cover a [3780-6910]~\AA$\;$ wavelength range, with a resolution of $\sim$120000. Table~\ref{tab_sample} lists these surveys and indicates the number of stars of our sample that were extracted from them. We note that  many stars have been studied by more than one group (hence the total is
larger than 360).

Temperatures are not available for all stars from a single source,
therefore we combined several references. As there is usually a systematic offset between these temperature scales (due to different methods -spectroscopic or photometric-, implementations, models, and data), we computed temperature shifts between pairs of samples (for stars in common) to apply a correction. As a large proportion  of our stars are in the sample studied by \cite{sousa08}, we used their temperature scale as a reference. We then used the following temperatures, in that order depending on availability: \cite{sousa08}, \cite{ramirez14} shifted by +5~K, \cite{gray06} shifted by +42~K, \cite{marsden14} shifted by +13~K, \cite{allende99} shifted by -41~K, and \cite{holmberg09} shifted by -71~K.

As for temperatures, ages were retrieved from various sources. The differences in age for a given star between different authors present a very large dispersion, which is a significant source of uncertainty on these values.
We used the following sources, in that order depending on availability: 
\cite{holmberg09}, \cite{delgado15}, \cite{ramirez14}, \cite{marsden14}, and \cite{borgniet15b}.

The activity level is characterized by the usual LogR'$_{\rm HK}$  (chromospheric emission), computed from the analysis of all spectra (see Paper I), as the LogR'$_{\rm HK}$ is strongly related to the presence of activity \cite[e.g.,][]{meunier10a}. It is therefore contemporary to the estimation of the convection amplitude performed in this paper.

Finally, the metallicities [Fe/H] were also extracted from different sources. We used the following sources, in this order depending on availability: \cite{sousa08}, \cite{ramirez14}, \cite{marsden14}, \cite{gray06}, \cite{holmberg09}, and \cite{delgado15}.  

Figure~\ref{sample} shows the distribution of these properties for all stars in our sample. The sample is biased toward old, solar-type, and not very active stars. 
We only included stars with a low $v$sin$i,$  that is, lower than 5~km/s, as in Paper I. The $v$sin$i$ values were mostly retrieved from \cite{nordstrom04}, and when not available there, from \cite{jenkins11}, \cite{valenti05}, \cite{dossantos16}, \cite{strassmeier00}, and \cite{lovis05}.  The reason for this selection is that in contrast to stars with a low $v$sin$i$, the  differential velocity shift for stars with large  $v$sin$i$ exhibits a significant decrease, which may be a bias. In principle, the velocity field (inside granules) affects different parts of spectral lines differently
because they correspond to different altitudes in the atmosphere. When the $v$sin$i$ is large, the contribution from different altitudes is mixed, which could affect the measured radial velocity, and we suggest that the differential velocity we determine (only
around line center) is mixed with the contribution from various altitudes. 


Table~\ref{tab_sample2} provides the list of the 360 stars in our sample as well as their properties and references. 

\subsection{Spectral line analysis}

\begin{figure} 
\includegraphics{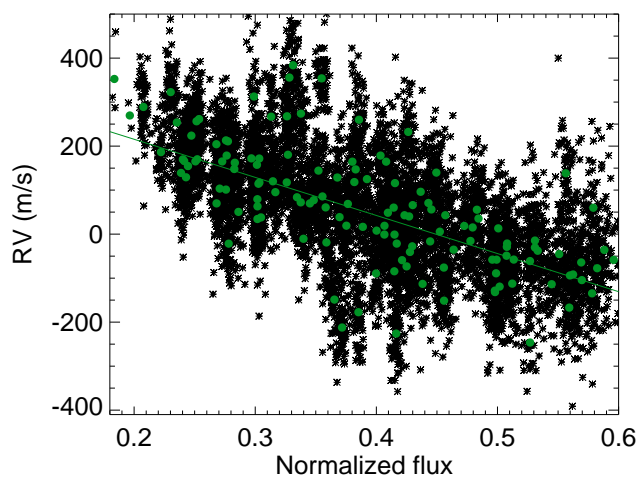}
\caption{
RV versus a normalized flux at the bottom of the lines for HD223171 (G2). Stars represent individual measurements, while the green dots correspond to the temporal average for each spectral line. The straight line is a linear fit on the green dots (from Meunier et al., 2017). 
}
\label{exemple}
\end{figure}

In this section, we briefly describe the analysis performed on the HARPS spectra for all stars in our sample. We refer to Paper I for more details and discussions. We took the following steps:

\begin{itemize}
\item{The continuum of the spectra is normalized to a flux of 1. A first estimation is made by determining the upper envelope, and it is followed by a correction by a factor taking into account the noise. The full procedure is detailed in Paper I. } 
\item{Spectral lines corresponding to TiI and FeII from \cite{dravins08} and FeI from \cite{nave94} are selected within the range 5000-6400 \AA. The position of the line is derived from a fit around line center for each of these lines on all individual spectra, and the velocity shift is computed from the difference between this wavelength and  the laboratory wavelength.  A total of 237 lines is used. We note that the convective blueshift is expected to be slightly higher \cite[][]{dravins86} for FeII lines than for FeI lines: we obtain convective blueshifts that remain within the observed dispersion for the TiI and FeI lines, and there are in fact very few FeII lines in our sample so that in practice our results is dominated by the TiI and FeI lines (as shown in the line list in Paper I). }\item{For each star and each observing time, we compute the averaged RV (i.e., over all spectral lines). This average is then subtracted from each individual velocity shift at that time step. We then average over time all velocities for a given spectral line. The resulting RV versus line flux for that star ( one per spectral line) is then fit with a linear function. }
\item{The slope of the linear fit measures the amplitude of the differential blueshift, which we named the TSS \cite[for "third signature slope", following the third signature proposed by][]{gray09},  in m/s/(F/Fc) (with F/Fc representing the intensity after normalization by the continuum Fc). We note that the unit m/s is used in the following to simplify the notations.   Individual and temporally averaged measurements are illustrated in Fig.~\ref{exemple}. 
}
\item{The TSS is used as a criterion to characterize the amplitude of the convective blueshift (i.e., the absolute blueshift) after a normalization with the solar values, based on the assumption made by \cite{gray09} that the shape of the differential shift of spectral lines is representative of the absolute convective blueshift: $RV_{\rm convbl} = TSS \times RV_{\rm convbl\odot} / TSS_{\odot}$. We normalized it to the solar values computed in Paper I: solar TSS of -776~m/s, which we derived from the solar spectrum of \cite{kurucz84} degraded to the HARPS spectral resolution; solar convective blueshift of 355 m/s, which we computed from the absolute solar RV versus spectral line depths of \cite{reiners16} for lines identified in the solar spectra of \cite{kurucz84} and assuming that the RV computed from cross-correlations between spectra will give more weight to deep lines. We recall that the derived absolute convective blueshift, either the solar value of 355 m/s or the stellar values inferred from the above relation, therefore correspond to this set of lines and procedure, and care must be taken when comparing with other works.} 
\end{itemize}

In addition, we computed the LogR'$_{\rm HK}$  from each spectrum. We use the average value for each star in Sect.~3 and consider its variability in Sect.~4. The RVs corresponding to our spectra computed by the Data Reduction Software (DRS) at ESO were also retrieved from the archive data to compare the two variabilities in Sect.~4. For some of these stars, we performed a correction when the variability was dominated by a binary component or a known planet.

\section{Differential velocity analysis}

\subsection{Teff, B-V, and activity relationship}

\begin{figure} 
\includegraphics{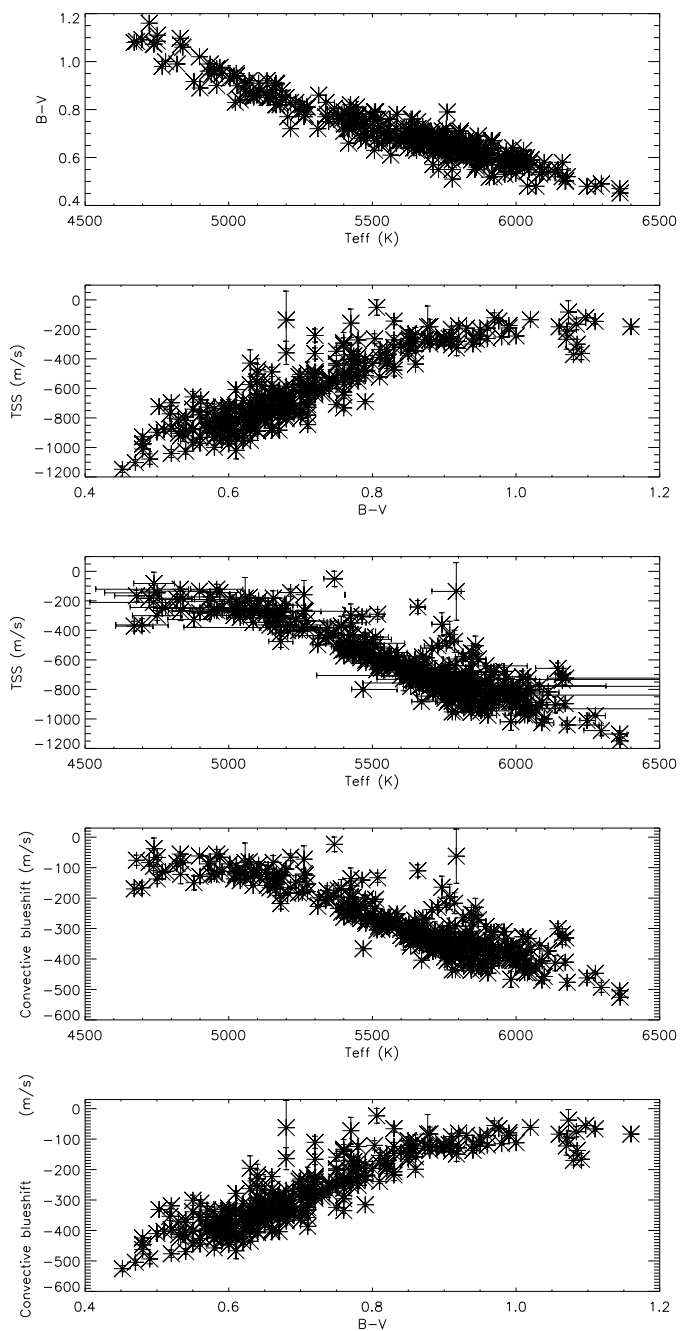}
\caption{
{\it First panel}: B-V versus Teff for the 360 stars in our sample.
{\it Second panel}: same for the TSS versus B-V.
{\it Third panel}: same for the TSS versus Teff.
{\it Fourth panel}: same for the convective blueshift versus Teff.
{\it Fifth panel}: same for the convective blueshift versus B-V.
}
\label{convbl}
\end{figure}

\begin{figure} 
\includegraphics{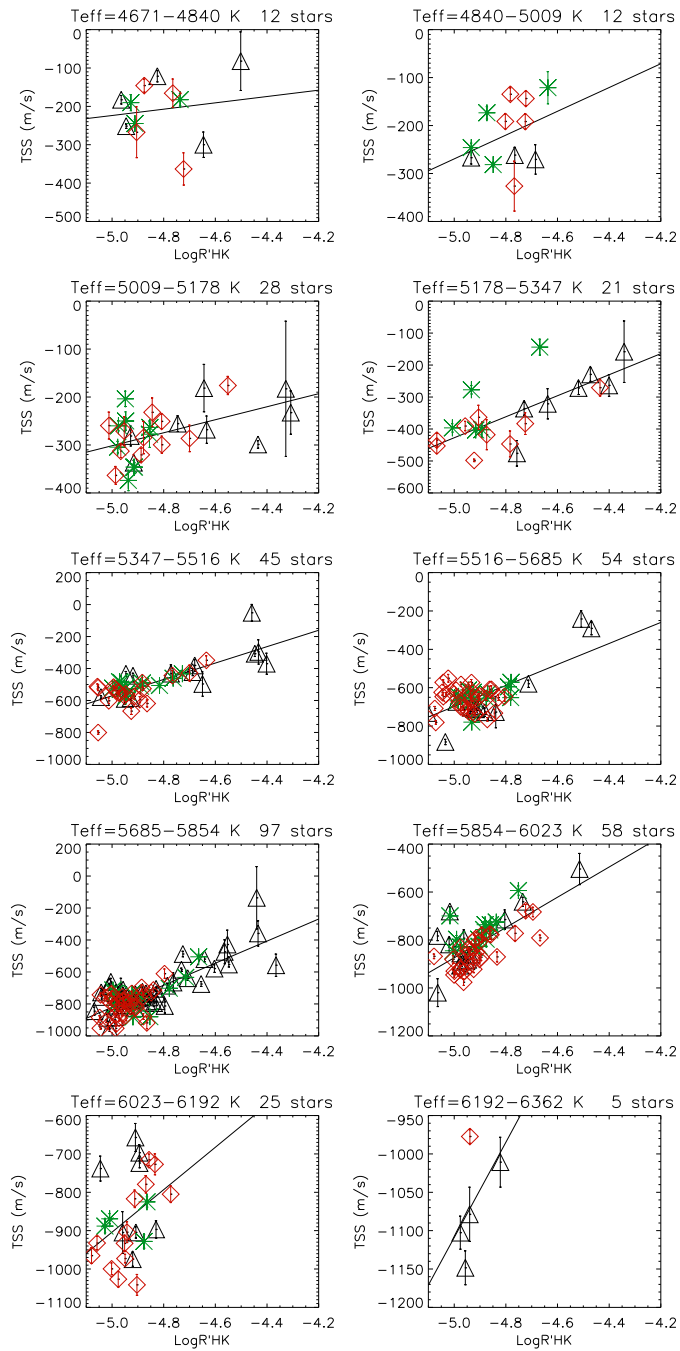}
\caption{
TSS versus LogR'$_{\rm HK}$  for ten Teff ranges. The solid line is a linear fit. Green stars are stars with an identified activity cycle, red diamond show stars with no cycle, and black triangles represent stars with undetermined behavior. }
\label{bte}
\end{figure}

\begin{figure} 
\includegraphics{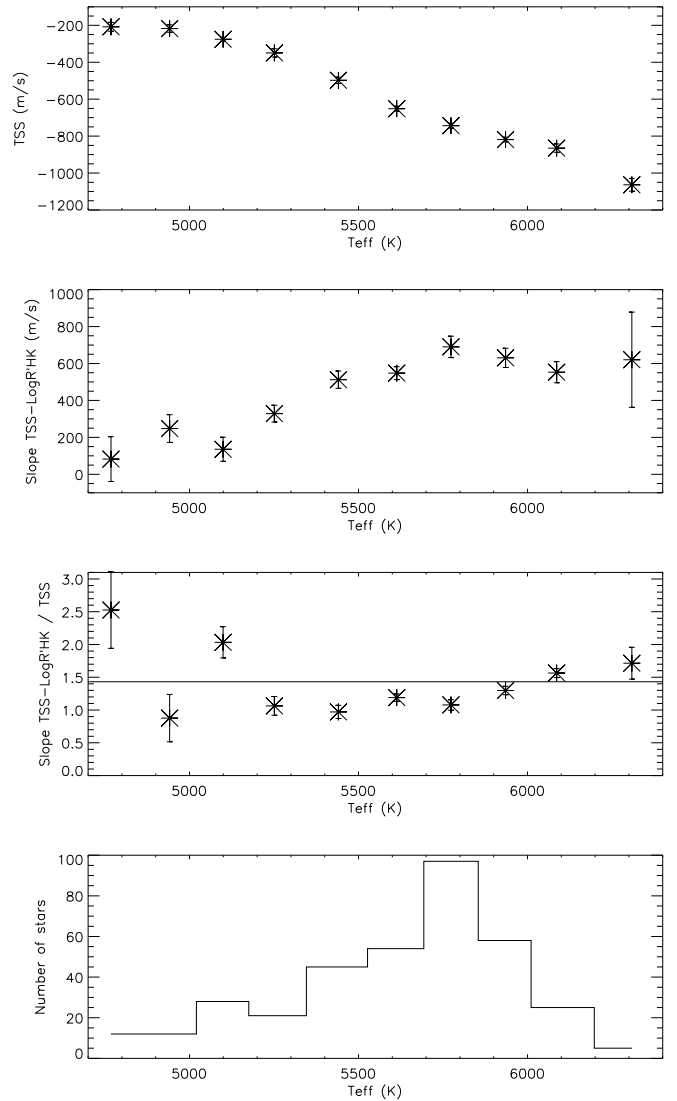}
\caption{
{\it First panel}: average TSS versus Teff in ten Teff bins.
{\it Second panel}: same for the slope of the TSS versus LogR'$_{\rm HK}$.
{\it Third panel}: same for the ratio between the slope and the average.
{\it Fourth panel}: number of stars in each bin.
}
\label{teff}
\end{figure}

The first panel of Fig~\ref{convbl} shows B-V versus Teff: there is a very good correlation, as expected, but we point out here the low dispersion in this relationship, to be compared to the other plots. 
The second and third panels show the TSS versus B-V and Teff ,respectively. We observe a very good relation as well, with a dispersion that is not much larger than the B-V versus Teff relation, and very few outliers. As pointed out in Sect.~2.1, there is a large uncertainty on the Teff values: the third panel shows the range covered by Teff from the different references indicated in Sect.~2.1 (we show it only in this panel for clarity). 
Finally, the last two panels show a reconstruction of the convective blueshift versus Teff and B-V, respectively, for our sample of 360 stars following the method described in Sect.~2.2. We do not see any saturation toward high Teff (or low B-V), as was marginally visible in Paper I. On the other hand, there is a clear saturation at low Teff (Teff below 5000 K), with a convective blueshift around 100~m/s.
As in Paper I, these convective blueshifts are higher than those derived from the numerical simulation made by \cite{allendeprieto13}.  Such a comparison is made after taking into account the difference in convective blueshift computed for our set of lines and the small wavelength range used by \cite{allendeprieto13} to simulate Gaia observations (i.e., 8470-8740~\AA). The spectral resolution and method to infer the wavelength shift are different, however, so that \cite{allendeprieto13} may have underestimated the effect as they used a cross-correlation between the spectra and a template spectrum to estimate the velocity shifts: a more precise analysis of simulated spectra should therefore be performed to understand the origin of the difference better.

Figure~\ref{bte} shows the individual TSS  (one average per star) versus LogR'$_{\rm HK}$  for ten Teff ranges covering the full available range. There is a good correlation between the two variables for almost all bins. A few correlations are not well defined, especially when combining a small sample (at very high or very low Teff) and a low amplitude of the TSS (at low Teff). 
The bin 5516-5685~K also lacks of stars at high activity levels.

Finally, Fig.~\ref{teff} shows the TSS averaged in each of these Teff bins versus Teff, as well as the slope of the TSS versus LogR'$_{\rm HK}$  and the ratio of the two. The last panel shows the number of stars in each bin, with a strong bias toward solar-type stars: some of the bins are not well sampled. The ratio between the slope and the average TSS is very interesting because it characterizes the attenuation of the convective blueshift that
is due to activity for a given variation of the activity level. The average is equal to 1.47 (1.27 when weighted by the inverse of the uncertainties to the power of two), with a 1$\sigma$ uncertainty of 0.18. 
Although there is no global trend over the full range, there is a trend for Teff higher than 5200 K, and the ratio can be modeled by -2.64+6.77 10$^{-4}$$\times$Teff: more massive stars tend to have a higher ratio than less massive stars. At a Teff of 6300 K, the attenuation should therefore be about 35\% stronger than for the Sun, while at a Teff of 5200K, the attenuation factor should be 25\% weaker than for the Sun. 
This means that the more massive stars, which also have a stronger convective blueshift, should exhibit even larger RV variations because of the attenuation of the convective blueshift in plages since the ratio is higher than for the Sun.



These observations on a much larger sample therefore confirm the results of Paper I, especially the global trend versus Teff and the dependence on activity, and extend the results to a wider range in Teff. The attenuation factor seems to exhbit a trend above 5200 K, which could not be detected from the smaller sample of Paper I.

\subsection{Effect of cyclic activity on the convection-activity relationship}

\begin{figure} 
\includegraphics{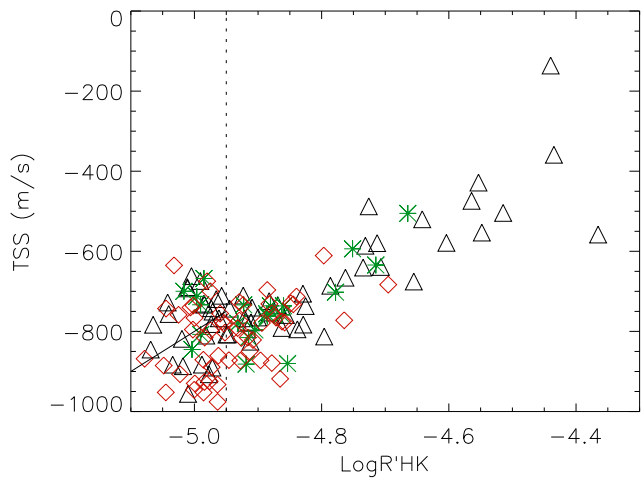}
\caption{
TSS versus LogR'$_{\rm HK}$  for the 137 stars in our sample with Teff between 5662~K and 5918~K. The color code is similar to that in Fig.~3. The solid line indicates the separation between the stars following the linear behavior and those with a lower convection level in Sect.~3.4, for stars with LogR'$_{\rm HK}$  lower than -4.95 (dotted line).
}
\label{cycle}
\end{figure}

Figure~\ref{cycle} shows the TSS versus LogR'$_{\rm HK}$  for a selection of stars with Teff in the range 5662-5918~K, selected to be similar to the range covered by the sample of \cite{marsden14}, who focused on solar-type stars. We identify 21 stars in this plot for which \cite{lovis11b} identified cyclic activity  (green) and 55 stars for which they identified strong variability, but no cycle (red). We find that these two populations do show a very similar behavior, within 2$\sigma$. It is a small sample,  however, especially for stars with a cycle, therefore there is a large uncertainty on the slopes. We note that for many stars in this Teff range, it is unclear whether they had a cycle, either because they were not studied by \cite{lovis11b} or because of poor sampling, which prevented such a characterization. A similar distinction in Fig.~\ref{bte} leads to a similar conclusion.

We observe in Fig.~\ref{cycle} a relatively large dispersion in TSS at low LogR'$_{\rm HK}$, with a deviation from the linear relationship toward a lower convection level. For LogR'$_{\rm HK}$  below -4.95, we have separated the sample (for Teff in the range 5662-5918~K) into two subsets: stars with a TSS above 1000$\times$LogR'$_{\rm HK}$ +4200 m/s (solid line in Fig.~\ref{cycle}), and stars with a TSS below this line. These two subsets do not show any significant difference in Teff, $v$sin$i$, or age. However, the first set has an average [Fe/H] of 0.13$\pm$0.02 and the other -0.09$\pm$0.02: this could therefore be due to a metallicity effect, which is studied in more detail in Sect.~3.4.

\subsection{Effect of age on the convection-activity relationship}

\begin{figure} 
\includegraphics{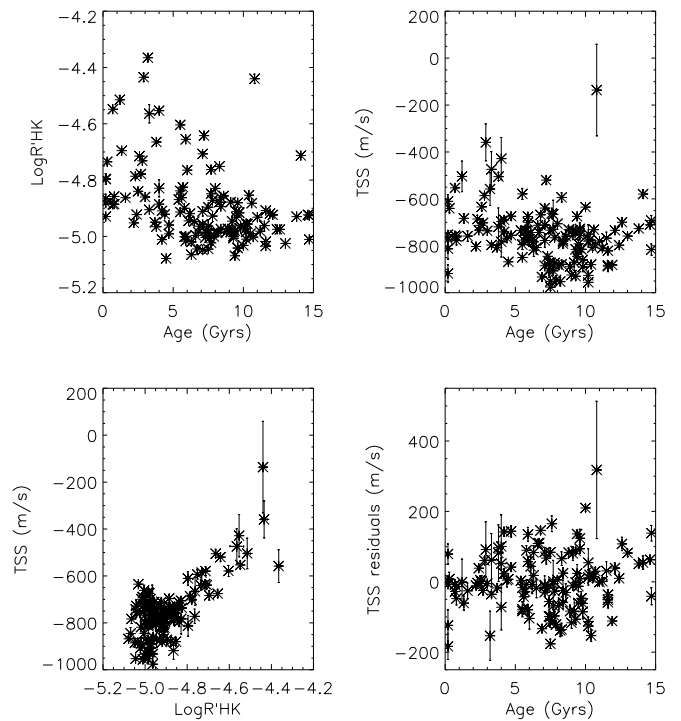}
\caption{
{\it Upper left panel}: LogR'$_{\rm HK}$  versus age for stars with Teff in the range 5662-5918~K.
{\it Upper right panel}: same for the TSS versus age.
{\it Lower left panel}: same for the TSS versus the LogR'$_{\rm HK}$.
{\it Lower right panel}: same for the residual TSS (after correction from the LogR'$_{\rm HK}$  trend) versus age.
}
\label{age}
\end{figure}

We now consider stars in the same temperature range, for which we have a relatively large homogeneous sample, and analyze the relationship between TSS and age. Ages are not identified for a few of these 137 stars, so we consider 133 stars here. Figure~\ref{age} shows the LogR'$_{\rm HK}$  and the TSS versus age, showing very similar behaviors. After correcting for the strong TSS-LogR'$_{\rm HK}$  dependence (shown in the lower left panel), the residuals versus age do not show any trend and are of much smaller amplitude. The age of the star therefore does not significantly affect the TSS, and it plays a role  only through the relationship between age and activity, which is the main factor. 

We also separated our sample into two subsamples: 13 stars with an age below 2 Gyr, and 120 older stars. We do not observe any difference in average TSS or in slope TSS versus LogR'$_{\rm HK}$ , within 1$\sigma$. The small sample of young stars limits this analysis, hosd df. When considering an age limit of 1 Gyr, the sample of young stars is limited  to 10 stars and the difference is even less significant.

We conclude that we do not see any significant difference in the behavior between the young and old stars in our sample, and estimate that plages, which are more frequently present in young stars (with a large LogR'$_{\rm HK}$ than in older stars) may have similar properties as plages in older stars.  

We note that the separation between young and old stars here
refers to their activity level. The age limit corresponding to the spot-dominated regime (young stars) and plage-dominated regime (old stars) in \cite{lockwood07} is not well defined, but our 2 Gyr threshold is consistent with the limit between spot-dominated and plage-dominated regimes in \cite{lockwood07}\footnote{The Hyades isochrone at 625 Myr \cite[][]{perryman98} falls well within the young star regions. The limit between the two regimes corresponds to older stars and  seems to exhibit a decreasing LogR'$_{\rm HK}$ as B-V increases, while we expect the isochrones to have the opposite behavior \cite[e.g.][]{mamajek08}: the age limit derived from the isochrones of \cite[e.g.][]{mamajek08} would then be close to 1 Gyr for B-V around 0.5, 2 Gyr around solar mass stars, and 3 Gyr for B-V around 1.}. 

We also recall that our young star subsample is biased toward slowly rotating stars because it is difficult to measure the TSS for stars with a $v$sin$i$ higher than 5 km/s. We therefore cannot exclude that plages in fast-rotating stars could have different properties. It may also be a bias toward stars seen pole-on.

\subsection{Effect of metallicity on the convection-activity relationship}

\begin{figure} 
\includegraphics{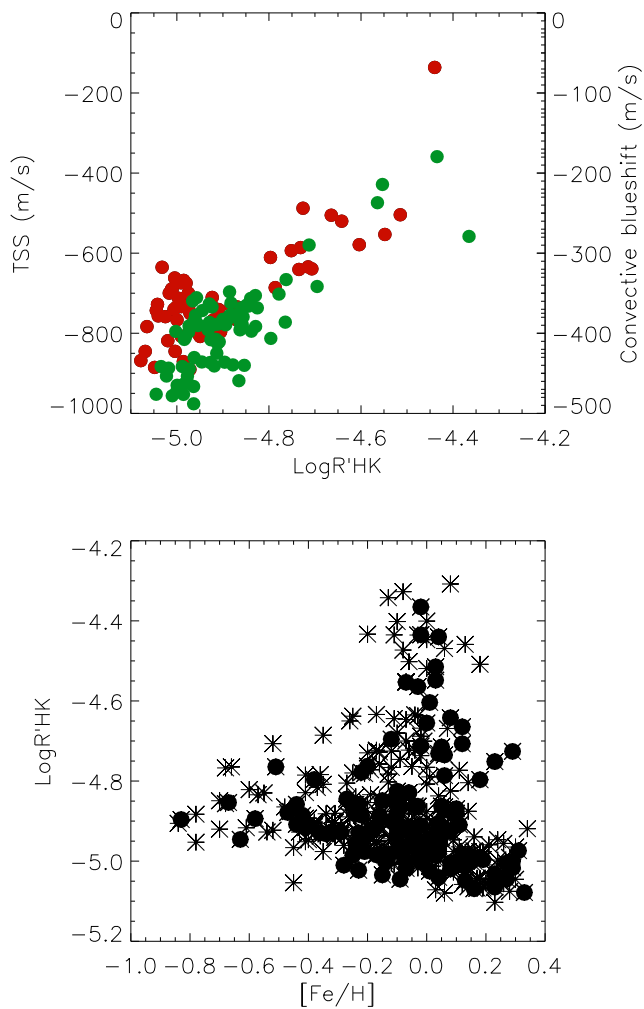}
\caption{
{\it Upper panel}: TSS (left axis) and convective blueshift (right axis) versus  LogR'$_{\rm HK}$ for Teff between 5662 and 5918~K, for stars with a positive [Fe/H] (red) and with a negative [Fe/H] (green). 
{\it Lower panel}: LogR'$_{\rm HK}$  versus [Fe/H] for stars with Teff 
between 5662 and 5918~K (circles) and the other stars of our sample (stars).
}
\label{met}
\end{figure}

We have shown in Sect.~3.2 that for stars in the 5662-5918~K range and LogR'$_{\rm HK}$  below -4.95, there is a strong link between the TSS and [Fe/H], the larger [Fe/H] being associated with a lower convective blueshift. We now investigate this effect in more detail. The upper panel of Fig.~\ref{met} shows separately the TSS versus LogR'$_{\rm HK}$  for Teff in the same range, but for different metallicities. The high-metallicity stars, showing the deviation toward a lower convective blueshift, also correspond to low LogR'$_{\rm HK}$  and a lower convective blueshift than the low-metallicity stars. 
Furthermore, we also observe that there is a relationship between the metallicity and the LogR'$_{\rm HK}$, as shown in the last panel: this plot is very similar to the results obtained by \cite{jenkins08}: for LogR'$_{\rm HK}$  smaller than -4.80, low metallicity is associated with more active stars. The origin of this effect is not discussed by \cite{jenkins08}, but it could be either due to the effect of metallicity on the estimation of the activity level using the LogR'$_{\rm HK}$ itself  \cite[the computation of
which does not include the effect of metallicity, see ][]{rochapinto98,wright04b,judge07,mittag13} , or it could be due to the fact that stars with a low metallicity are more active, for example because they have a more vigourous small-scale convection (although we found no suggestion of this possibility in the literature), leading to a higher convective blueshift. 

Our interpretation of these plots is the following: we observe two competing effects here. On one hand, low metallicity seems to be associated with a higher convective blueshift. On the other hand, it is also associated with higher activity levels, hence a reduced convective blueshift. This can explain the position of the high-metallicity stars in Fig.~\ref{met}, with a lower |TSS| than would be expected from the linear law.


We have quantified this difference in Appendix B and found that for a $\Delta$[Fe/H] of 0.32, the convective blueshift is indeed 10-12\% lower (in absolute value) for stars with the highest metallicity compared to those with the lowest.  

Several groups have studied the effect of metallicity on convection using hydrodynamical simulations of convection in various types of stars, with complex effects. \cite{allendeprieto13} found more vigourous convection for low metallicities (as observed here), although to a lesser degree: for Teff=5700 K and log g=4.5, the difference in velocity shift (with the factor used in this paper) between a $\Delta$[Fe/H] of 0.27 is 4.4 m/s, which represents 2.4\% of the low-metallicity value, that is, it is 4-6 times lower than our estimation. \cite{magic13} also found a higher granulation contrast for low-metallicity stars (around 25\%), but surprisingly, this did not seem to produce a significant difference in velocity shift for the Teff corresponding to our sample \cite[][]{magic14}. \cite{tremblay13} also found a higher granulation contrast at low metallicity, but with a smaller effect of metallicity than \cite{magic13}. Therefore, although simulations are qualitatively in agreement with our observations (low metallicity corresponding to a more vigourous small-scale convection), the observed effect of metallicity seems to be significantly larger than what was obtained in numerical simulations.  

Metallicity is important because some results have indicated that stars hosting giant planets may be more metallic on average than the Sun, leading to some conclusion on formation processes \cite[e.g.,][]{santos03,fischer05}. 
Exoplanet surveys are usually biased toward less active stars \cite[e.g.,][]{lovis11b}, however, to limit the effect of activity on RV, based on LogR'$_{\rm HK}$  values, but if low LogR'$_{\rm HK}$  values are associated with higher metallicities, as shown by \cite{jenkins08}, then this could alter the metallicity-exoplanet relationship. This may explain why such a relationship is not observed in transit surveys  \cite[][]{fridlund10}.

\section{Relationship between activity and RV variations}

\subsection{Statistical analysis of the sample}

\begin{table}
\caption{Statistics of the RV-LogR'$_{\rm HK}$  relationship}
\label{tab_stat}
\begin{center}
\renewcommand{\footnoterule}{}  
\begin{tabular}{llll}
\hline
Categ.     &  Definition                        &   NB  &  Perc.  \\ 
   &   &   &   (\%)   \\ \hline
1      &  Good sampling, excellent correlation &   13   &    7.9       \\2      &  Moderate sampling, good correlation&     48   &    29.0       \\
3      &  Weak correlation                    &    15   &    9.1       \\4      &  Weak signals, no correlation       &     70   &    42.4       \\
5      &  Strong LogR'$_{\rm HK}$, weak RV            &     19   &    11.5       \\
1-5 &                                   &     165   &     -     \\ \hline6      &  Strong signals, no correlation     &     12   &    28.6      \\7      &  Strong signals, anticorrelation     &   11   &    26.2     \\
8      &  Strong RV, weak LogR'$_{\rm HK}$              &    19   &    45.2      \\
6-8 &                                      &   42   &      -   \\
\hline
9   & Strong RV, identified origin       & 26 & - \\
\hline
\end{tabular}
\end{center}
\tablefoot{Number of stars in each category NB, for the selected stars (207).
The percentages indicate the fraction with respect to 148 for categories 1-5 and 42 for categories 6-8. Category 9 corresponds to a strong RV signal with a known origin (mostly presence of planets). Correlation means between RV and LogR'$_{\rm HK}$  time series; weak-strong signals are related to the {\it amplitudes} of RV and logR'$_{\rm HK}$, especially on long or median timescales, and not to the average level. }
\end{table}
%

\cite{meunier10a} have shown that long-term RV variations should be dominated by the inhibition of the convective blueshift in solar-type stars, if they behave like the Sun. There are indications of correlations between RV amplitudes and LogR'$_{\rm HK}$  amplitudes, as shown by the trend of the RV-LogR'$_{\rm HK}$  slope versus Teff obtained by \cite{lovis11b}. \cite{isaacson10} also estimated the RV jitter versus average activity level for different stellar types, but the derived lower envelope as a function of spectral type is not well constrained. In Paper I, in addition to characterizing the convective blueshift - activity relationship, we have also estimated the amplitude in RV compared to the amplitude in LogR'$_{\rm HK}$  for a small sample of stars and observed a trend, with a large dispersion. Overall, it is not clear up to which point solar-type stars behave like the Sun as obtained by \cite{meunier10a}. With a larger sample such as studied in this paper, we now have the opportunity to estimate the proportion of stars that follows what we call the "solar pattern" (RV long-term variations correlated with activity variability due to the inhibition of the convective blueshift, with a possible modulation of the effect with inclination, angle between the rotation axis, and the line of sight, and including small variability for both RV and LogR'$_{\rm HK}$), and those that do not, keeping in mind possible biases in our sample. We therefore now characterize the stars in our sample, following the characteristics  described in Table~\ref{tab_stat}. 

We first eliminate some stars from this statistical analysis for several reasons, mostly because they do not have enough points to characterize any long-term variations, and a few because the $RV$ time series from the ESO DRS were not reliable. We also counted 26 stars (category 9 in Table~\ref{tab_stat}) separately with known sources of $RV$ variations that we did not correct for: 4 binary stars, 2 variable stars, and 20 stars with a known planetary signal that we did not correct.

We identified 165 stars that follow the solar pattern: their  RV and LogR'$_{\rm HK}$  are correlated, or if they are not, both exhibit a small variability. We also included the  stars with large LogR'$_{\rm HK}$  variation but small RV variation, which could be due to a low convection level or inclination effects \cite{borgniet15}, or both. These 165 stars are categorized into five groups (categories 1-5)  in Table~\ref{tab_stat}. 

A second group of 42 stars were identified and did not follow this solar pattern, as they exhibit large RV variations that cannot be explained by their LogR'$_{\rm HK}$  variations: we observed either no correlation, an anticorrelation, or a weak LogR'$_{\rm HK}$  variation (categories 6 to 8). This means that for these stars the RV variations are dominated by another unknown effect. When we add the 26 stars mentioned above whose RV variations are dominated by a known effect other than the convection inhibition, this represents 29\% of this sample of 233 stars (18\% with unknown
origin, 11\% with a known origin), while those following the pattern represent 71\%. 
 
The definition of the categories is such that there is a certain continuum between them, which introduces an uncertainty, probably on the order of a few percent. We also recall that the sample is biased toward solar-type stars, for which we have shown that
the inhibition of the convective blueshift is important, so we may underestimate category 7 (unless lower mass stars have a larger variability, which can be observed from the analysis of the cycle amplitudes as a function of the average activity level in Lovis et al. 2011). On the other hand, it is also biased toward stars that are not very active (most are less variable than the Sun), therefore we may underestimate categories 1,2, and 8. In this small sample we do not observe any significant difference between the number of stars in the different categories for low- and high-mass stars, however. 


\begin{table}
\caption{Young star categories}
\label{tab_young}
\begin{center}
\renewcommand{\footnoterule}{}  
\begin{tabular}{llll}
\hline
Categ. & NB  & NB & Fraction of  \\ 
      &  young stars & total & young stars \\ \hline
1-5 & 13 & 165 & 0.079$^{+0.030}_{-0.022}$ \\
6-8 & 13 & 42 & 0.309$^{+0.158}_{-0.116}$ \\
9 & 4 & 26 & 0.154$^{+0.133}_{-0.089}$ \\
6-9 & 19 & 68 & 0.279$^{+0.111}_{-0.087}$ \\
\hline
\end{tabular}
\end{center}
\tablefoot{Categories are defined in Table 2. Young stars are those younger than 2 Gyr. }
\end{table}

Finally, we examined the distribution of young stars in these categories. The results are summarized in Table.~\ref{tab_young}. Young stars seem to be overrepresented in categories 6-8 or 6-9, which is significant at the 1$\sigma$ level.  

\subsection{RV variability}

\begin{figure} 
\includegraphics{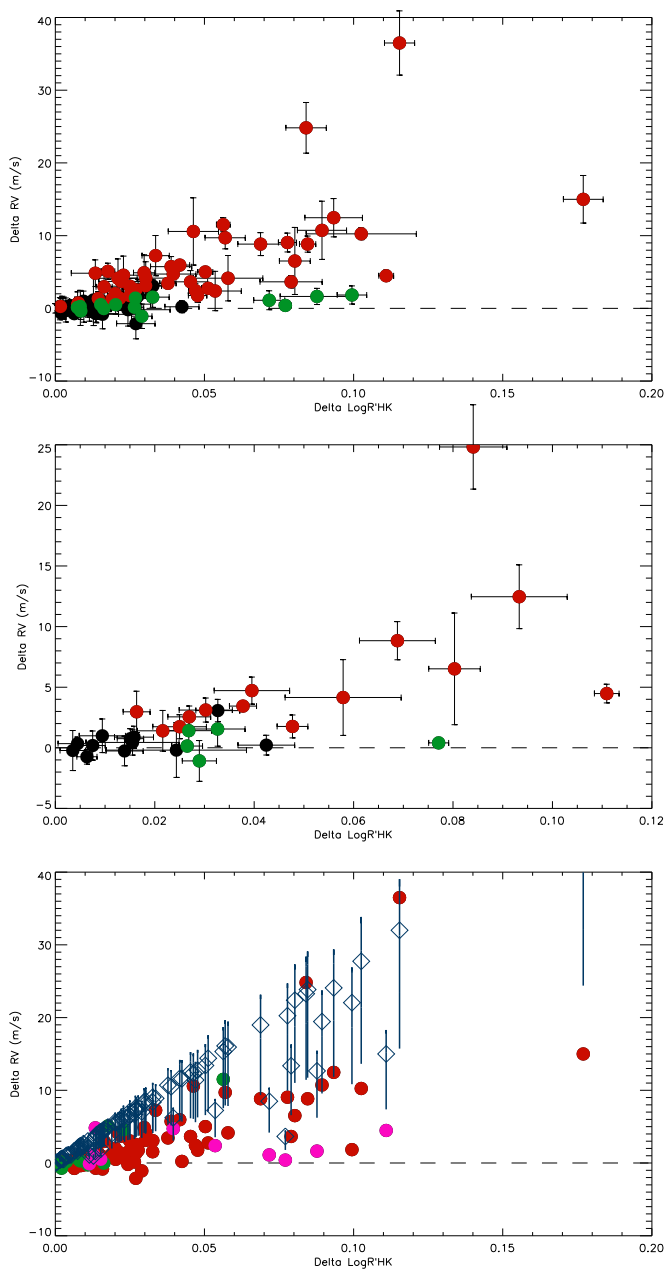}
\caption{
{\it Upper panel}: $\Delta$ RV versus $\Delta$LogR'$_{\rm HK}$  for our selection of 106 stars, from categories 1-3 (red), category 5 (green), and category 4 (black).
{\it Middle panel}: same for the 31 stars with $\Delta$t larger than 1000 days. 
{\it Lower panel}: same as the upper panel, but showing the different spectral types: F stars (green circles), G stars (red circles), and K stars (pink circles). The blue diamonds indicate the reconstructed RV amplitude, and each vertical blue line shows the expected effect of inclination. 
}
\label{delta}
\end{figure}

We now focus on stars that follow the convection inhibition pattern, that is, stars in categories 1-5. As in Paper I, we selected stars with enough points to estimate a reasonably representative long-term amplitude and average the values in bins of at least five points. We identified the observing times corresponding to the minimum and maximum LogR'$_{\rm HK}$, for these two observing times we computed $\Delta$LogR'$_{\rm HK}$ and $\Delta$RV. This led to a selection of 106 stars. We introduced an additional selection on the time range $\Delta$t separating the two observing times identified as minimum and maximum: we then considered a second subsample of stars for which $\Delta$t is larger than 1000 days as well (to focus on long-term variations), which led to a sample of 31 stars.

The results are shown in Fig.~\ref{delta} for these two selections of 106 and 31 stars. We do observe a trend with larger $\Delta$LogR'$_{\rm HK}$  for larger $\Delta$RV. As in Paper I, there is a large dispersion, which was discussed as being due not only to the expected difference in response for different spectral types, but also to inclination effects that cannot be removed. We also note that although the sample has been improved, there is a lack of stars with a large $\Delta$LogR'$_{\rm HK}$: there are almost no stars in our sample with an amplitude larger than solar (0.1 at least), which is a problem. 

We note a few outliers with $\Delta$RV between -10 and -5 m/s. This is usually within the noise at the 3$\sigma$ level and could be due to particular sampling effects. We also recall that there is a certain continuity between the categories, so these few stars could be close to categories 6-9. 

As in Paper I, we also estimated for each star the RV amplitude $\Delta$RV$_{\rm conv}$ that is expected given their observed $\Delta$LogR'$_{\rm HK}$, assuming they follow the laws observed in Sect.~3 and that the RV amplitude is only due to the inhibition of the convective blueshift (see Paper I for more details):

\begin{equation}
\Delta RV_{\rm conv} = G \times  RV_{\rm convbl \odot} / {\rm TSS_{\rm \odot}} \Delta LogR'_{\rm HK}
,\end{equation}
where $G$ is the slope of the TSS versus LogR'$_{\rm HK}$  for the considered Teff (obtained using a polynomial fit on the curve shown in Fig.~\ref{teff}, second panel), RV$_{\rm convbl \odot}$ and TSS$_{\rm \odot}$ are the solar values. 
We also took into account the fact that stellar inclinations are expected to significantly affect the observed $\Delta$RV$_{\rm conv}$. There is a general agreement in the trend with observations. The reconstructed values tend to be on average slightly higher than the observed values, although for most points this is within the 3$\sigma$ uncertainties. There are very few outliers above, which means that our star selection is good (we did not include stars with RV variations that cannot be explained by the logR'$_{\rm HK}$  behavior). 

However, we still note a few stars for which we would have expected a larger RV amplitude given their convection level and activity variability. The most noticeable is HD196390 (G1, $\Delta$LogR'$_{\rm HK}$ =0.18, lower right corner of the plot), which we classified into category 2: it shows a very well defined and large-amplitude activity variation, and the RV variation, although correlated (correlation factor of 0.58), is indeed noisy. The slope of a linear fit of the RV versus LogR'$_{\rm HK}$  multiplied by the observed $\Delta$LogR'$_{\rm HK}$  leads to an RV amplitude of 15~m/s, which is closer to the expected value. This illustrates the limits of the analysis and the effect of the sampling on the estimations of $\Delta$LogR'$_{\rm HK}$  and $\Delta$RV, in addition to the intrinsic dispersion that is due to inclination. We therefore also tested another estimation of the observed $\Delta$RV by multiplying the slope RV versus LogR'$_{\rm HK}$  for each star with $\Delta$LogR'$_{\rm HK}$: this gives similar results, although there are fewer outliers (such as HD196390 or stars with a negative $\Delta$RV). The comparison between observed and predicted $\Delta$RV therefore remains difficult. 

The effect of inclination mentioned above refers to the difference in long-term amplitude of the variations of LogR'$_{\rm HK}$ and $RV_{\rm conv}$ when a star like the Sun is seen from different angles, as shown by \cite{borgniet15}: Because the activity belt is located close to the equator, both amplitudes decrease when we move from a view from the equator to a pole-on point of view for a solar-type activity pattern. However, the long-term amplitude of $RV_{\rm conv}$ decreases faster because it cumulates two projection effects: the apparent size of the structure decreases toward the limb, and the projection of the convective blueshift along the line of sight decreases as well. In principle, another effect could be taken into account, that is, the effect of a variable convective blueshift with latitude (independently of magnetic activity), which is neglected here: solar granulation does not show strong variability with latitude, however \cite[][]{rodriguez92}, and any variation seems to be mostly associated with magnetic structures (taken into account in the simulation). 
Therefore we do not expect the convective blueshift itself to vary significantly with inclination.
We also note that we used a single law describing the relationship between activity level and TSS, which corresponds to the average inclination in our sample. Stars with similar average activity level but different inclinations should appear at a different position in the plots of Fig.~\ref{bte} for example: compared to a pole-on observation, a star with a different inclination will appear at a lower activity level (if the activity belt is close to the equator, as for the Sun) and a higher |TSS| value as well, possibly with a proportionaly weaker attenuation given the projection effects. It is therefore possible that the TSS-LogR'$_{\rm HK}$  laws exhibit slightly different slopes depending on the inclination, and they therefore contribute to the observed dispersion.  

Finally, we note that our sample includes mostly G stars, and as shown in the figure, there are very few K stars. We find that they lie below the G stars for a given logR'$_{\rm HK}$ (orange circles in the lower panel of Fig.~\ref{delta}), which is expected. The F stars lie at the top, but not above, the G stars, showing that the flattened part at high Teff in the slope versus activity (Fig.~\ref{teff}, second panel) may be real.

\section{Conclusion}


We have analyzed a sample of 360 main-sequence stars with Teff between 4600 and 6400 K (K4 to F7), biased toward old stars, but also including a few relatively young very active stars (younger than 2 Gyr), and derived the amplitude of the convective blueshift for each star using the differential shift of spectral lines as in Paper I. We obtained the following results:

\begin{itemize}
\item{We confirm the strong variation in convective blueshift with Teff obtained in Paper I. We observe a saturation toward low-mass stars. }\item{We confirm the dependence on the activity level. The attenuation factor of convection with activity seems to increase with Teff for Teff in the range 5200-6300~K, although it is not possible to identify any trend for Teff  below this range. In the smaller sample of Paper I, no trend was identified, which was in good agreement with the numerical simulation of \cite{steiner14} and \cite{beeck15}, showing a similar magnetic field strength in magnetic structures for the different spectral types. If the magnetic field remains constant, it remains to be understood why the attenuation would be more efficient for more massive stars.
}
\item{We observe no significant difference between stars with a cycle and those without a cycle, suggesting similar properties of their active regions.}
\item{Similarly, we have not been able to identify a significant difference between the youngest stars in our sample (younger
than 2~Gyr) and the other stars, also suggesting that plages are not very different.}
\item{On the other hand, we observe a significant effect of metallicity, at least for solar-type stars for which we have a large sample: we estimate that stars with a high  metallicity exhibit a lower convection level (as measured by the TSS and the convective blueshift) than those with a low metallicity, on the order of 10\% for a $\Delta$[Fe/H] of 0.27, which is at least four times higher than predicted by numerical simulations. We also confirmed the dependence of the LogR'$_{\rm HK}$ on metallicity observed by \cite{jenkins08} and discussed the possible effect on the possible relationship between metallicity and exoplanet that is due to biases in exoplanet surveys. }
\item{We computed a percentage of 71\% of stars that follow the "solar pattern", that is, stars that exhibit a good correlation between activity and RV variability, including stars with no variability for either variables, or stars with a low RV variability but a high LogR'$_{\rm HK}$ variability (which could be due to inclination effects). Keeping in mind that the sample might be biased, we estimate that three quarters of the stars follow this pattern. A quarter of the stars exhibits high RV variations,
however, which cannot be attributed to the inhibition of convection and which are probably due to other processes. }
\end{itemize}

\begin{acknowledgements}

This work has been funded by the Universit\'e Grenoble Alpes project call "Alpes Grenoble Innovation Recherche (AGIR)" and on the ANR GIPSE ANR-14-CE33-0018.
This work made use of several public archives and databases: 
The HARPS data have been retrieved from the ESO archive at http://archive.eso.org/wdb/wdb/adp/phase3\_spectral/form.
This research has made use of the SIMBAD database, operated at the CDS, Strasbourg, France. 
Exoplanet information
has been retrieved from the Extrasolar Planet Encyclopaedia at http://exoplanet.eu/. The solar spectrum 
from Kurucz et al. (1984) is available online at  
http://kurucz.harvard.edu/sun/fluxatlas2005/. 

\end{acknowledgements}
\bibliographystyle{aa}
\bibliography{bib31017}

%

\begin{appendix}

\section{Additional tables}

\onecolumn
\onllongtab{1}{
\begin{longtable}{lllllllllllll}
\caption{\label{tab_sample2} Star properties}\\
\hline
Name & Teff  & B-V & Sp. T. & N$_{\rm Spectra}$ & TSS & $\sigma$TSS & LogR'$_{\rm HK}$ & $\sigma$ & Conv.  & Age & $v$sin$i$ & Source \\
    &         &     &       &                   &  (m/s   &  (m/s           & & LogR'$_{\rm HK}$ & Blueshift & & & \\
    &         &     &  (K) &            & /(F/Fc))&  /(F/Fc))     &        &  & (m/s)  & Gyrs & km/s &   \\
\hline
HD169830 & 6361$^{1}$ & F7 & 0.47 &    65 & -1102.6 &   21.5 & -4.976 &  0.003
 &  -504.4 & 2.30$^{10}$ & 4$^{13}$ & 1\\
HD184985 & 6362$^{8}$ & F7 & 0.45 &    38 & -1148.4 &   22.1 & -4.957 &  0.002
 &  -525.4 & 2.40$^{10}$ & 5$^{13}$ & 5\\
HD693 & 6297$^{8}$ & F8 & 0.49 &    35 & -1078.7 &   35.3 & -4.940 &  0.002 & 
 -493.5 & 3.80$^{10}$ & 5$^{13}$ & 5\\
HD11226 & 6098$^{1}$ & F8 & 0.58 &    38 &  -999.6 &   15.4 & -5.002 &  0.004 & 
 -457.3 & 4.20$^{10}$ & 3$^{13}$ & 1\\
HD38382 & 6082$^{1}$ & F8 & 0.53 &    29 &  -904.6 &   14.9 & -4.908 &  0.004 & 
 -413.8 & 2.10$^{10}$ & 3$^{13}$ & 1\\
\hline
\end{longtable}
\tablefoot{This table is published in its entirety at the CDS. A portion is shown here for guidance regarding its form and content. Star name, spectral type and B-V (both from the CDS), Teff, number of spectra used in the analysis,
TSS and its 1$\sigma$ uncertainty, averaged LogR'$_{\rm HK}$ and its 1$\sigma$ uncertainty, convective blueshift derived from the TSS, age, $v$sin$i,$ and origin (survey) of the observations. The origins of the observations (references 1 to 7) are defined in Table~\ref{tab_sample}: 1 \cite[][]{sousa08}, 2 \cite[][]{ramirez14}, 3 \cite[][]{marsden14}, 4 \cite[][]{datson14}, 5 \cite[][]{borgniet17}, 6 \cite[][]{lagrange13}, and 7 \cite[][]{gray15}. The sources for Teff, age, and $v$sin$i$ are either from these references (1-7) or from 8 \cite[][]{gray06}, 9 \cite[][]{allende99}, 10 \cite[][]{holmberg09}, 11 \cite[][]{delgado15}, 12 \cite[][]{borgniet15b}, 13 \cite[][]{nordstrom04}, 14 \cite[][]{jenkins11}, 15 \cite[][]{valenti05}, 16 \cite[][]{dossantos16}, 17 \cite[][]{strassmeier00}, and 18 \cite[][]{lovis05}. 
}
}

\section{Metallicity effect}

To quantify the difference in TSS for stars of different metallicity as shown in Sect.~3.4 (see Fig.~\ref{met}), we computed the linear law fitting the TSS versus LogR'$_{\rm HK}$  only for stars with a negative metallicity (average [Fe/H] of -0.19, or -0.13 when restricted to logR'$_{\rm HK}$  lower than -4.95). Then, below -4.95, we estimated the average TSS that stars with a positive metallicity would have if they were following this linear law, compared to the observed value: this gives -754 m/s and -877 m/s, respectively. The same procedure was applied to the convective blueshift and  gives -401 and -345 m/s respectively, hence a difference of 56 m/s (14\% with respect to the linear law) for a $\Delta$[Fe/H] of 0.27. The effect is similar in the Teff range 5300-5662~K, with a difference of 38 m/s on the convective blueshift representing 12\% for a $\Delta$[Fe/H] of 0.32.

When we assume that for logR'$_{\rm HK}$  below -4.80, the observed trend of LogR'$_{\rm HK}$  versus [Fe/H] is entirely due to the effect of the metallicity on the LogR'$_{\rm HK}$  measurement (and not an intrinsic variation), we can correct  the LogR'$_{\rm HK}$ values for this trend and again apply the same procedure. For the Teff range 5662-5918 K, this gives a convective blueshift of -345 m/s (average) and -385 (from the negative metallicity linear fit), hence a difference of 40 m/s representing 10\%: this is slightly smaller than before correction, but on the same order of magnitude, so this result seems to be robust. We also note that \cite{rochapinto98} studied the difference between the chromospheric and isochrone ages, which would in principle also provide a way to correct for the effect of metallicity on LogR'$_{\rm HK}$  measurements, but the resulting correction seems unrealistic, which could be due to the very  large uncertainties on age estimations. 

\end{appendix}

\end{document}